\documentclass[runningheads]{llncs}
\pdfoutput=1

\usepackage{times}
\usepackage{helvet}
\usepackage{courier}

\def\beq{\begin{equation}}
\def\eeq#1{\label{#1}\end{equation}}
\def\ba{\begin{array}}
\def\ea{\end{array}}

\def\i#1{\hbox{\it #1\/}}
\def\ar{\leftarrow}

\def\rar{\rightarrow}
\def\lrar{\leftrightarrow}

\def\pnn{\hbox{\rm Pnn}}

\def\nnn{\hbox{\rm Nnn}}

\def\sm{\hbox{\rm SM}}

\def\comp{\hbox{\rm Comp}}

\def\bx{\hbox{\bf{x}}}
\def\by{\hbox{\bf{y}}}

\def\bp{\hbox{\bf{p}}}

\def\bt{\hbox{\rm\bf t}}

\def\uc{\widetilde\forall}

\def\gr{\i{gr}}

\usepackage{amssymb}

\usepackage{latexsym}
\usepackage{graphicx}
\usepackage{graphics}
\usepackage{amsmath}
\usepackage{longtable}
\usepackage{multirow}
\usepackage{verbatim}
\usepackage{xspace}
\usepackage{algorithm}
\usepackage{algorithmic}
\usepackage{epsfig}
\usepackage{array}
\usepackage{amssymb} % for math
\usepackage{amstext} % for text in math forumla

\title{\bf Lloyd-Topor Completion
and General Stable Models}
\author{Vladimir Lifschitz and Fangkai Yang}
\institute{Department of Computer
Science\\ The University of Texas at Austin \\
\{vl,fkyang\}@cs.utexas.edu}
\titlerunning{Lloyd-Topor Completion and General Stable Models}
\authorrunning{V.~Lifschitz and F.~Yang}

\begin{document}

\setcounter{page}{145}

\maketitle
\begin{abstract}
We investigate the relationship between the generalization of program
completion defined in 1984 by Lloyd and Topor and the generalization
of the stable model semantics introduced recently by Ferraris {\em et al.}
The main theorem can be used to characterize, in some cases, the general
stable models of a logic program by a first-order formula.  The proof uses
Truszczynski's stable model semantics of infinitary propositional formulas.
\end{abstract}

\section{Introduction}

The theorem by Fran\c{c}ois Fages \cite{fag94} describing a case when the
stable model semantics
is equivalent to program completion is one of the most important
results in
the theory of stable models.  It was generalized
in~\cite{lif96b,erd03,lin03},
it has led to the invention of loop formulas \cite{lin04}, and it has
had a
significant impact on the design of answer set solvers.

The general stable model semantics defined in \cite{fer09}
characterizes the
stable models of a first-order sentence~$F$ as arbitrary models of a
certain
second-order sentence, denoted by $\sm[F]$;\footnote{To be precise,
the
definition of $\sm$ in that paper requires that a set of ``intensional
predicates'' be specified.  In the examples below, we assume that all
predicate
symbols occurring in~$F$ are intensional.}
logic programs are viewed there as
first-order sentences written in ``logic programming notation.''
In this note we define an extension of Fages' theorem that can be used
as a
tool for transforming $\sm[F]$, in some cases, into an equivalent
first-order
formula.  That extension refers to the version of program completion
introduced by John Lloyd and Rodney Topor in \cite{llo84a}.  Their definition
allows the body of a rule to contain
propositional connectives and quantifiers.

Earlier work in this direction is reported in \cite{fer09} and
\cite{lee11b}.  These papers do not mention completion in the sense of
Lloyd and Topor explicitly.  Instead, they discuss ways to convert a
logic program to ``Clark normal form'' by strongly equivalent
transformations~\cite{lif01,lif07a} and completing programs in this
normal
form by replacing implications with equivalences.  But this is
essentially
what Lloyd-Topor completion does.

The following examples illustrate some of the issues involved.
Let~$F$ be
the program
\beq
\ba l
p(a),\\
q(b),\\
p(x)\ar q(x),
\ea
\eeq{a1}
or, in other words, the sentence
$$p(a)\land q(b)\land\forall x(q(x)\rar p(x)).$$
The Clark normal form of~(\ref{a1}) is tight in the sense of
\cite{fer09}, and Theorem~11 from that paper shows
that $\sm[F]$ in this case is
equivalent to the conjunction of the completed definitions of~$p$ and~$q$:
\beq
\ba l
\forall x(p(x)\lrar x=a \lor q(x)),\\
\forall x(q(x)\lrar x=b).
\ea
\eeq{a1comp}

Let now~$F$ be the program
\beq
\ba l
p(x)\ar q(x),\\
q(a)\ar p(b).
\ea
\eeq{a2}
This program is not tight in the sense of \cite{fer09}, so that
the above-mentioned theorem is not applicable.  In fact, $\sm[F]$ is
stronger in this case than the conjunction of the completed definitions
\beq
\ba l
\forall x(p(x)\lrar q(x)),\\
\forall x(q(x)\lrar x=a\land p(b)).
\ea
\eeq{a2comp}
A counterexample is provided by any
interpretation that treats each of the symbols~$p$,~$q$ as a singleton
such that its element is equal to both~$a$ and~$b$.  Such a (non-Herbrand)
interpretation
satisfies~(\ref{a2comp}), but it is not a stable model of~(\ref{a2}).
(In
stable models of~(\ref{a2}) both~$p$ and~$q$ are empty.)

Program~(\ref{a2}) is, however, atomic-tight in the sense
of~\cite[Section~5.1.1]{lee11b}.  Corollary~5 from that paper allows
us to
conclude that
the equivalence between $\sm[F]$ and~(\ref{a2comp}) is entailed by the
unique name assumption~$a\neq b$.
It follows that the result of applying~$\sm$ to the program obtained
from~(\ref{a2comp}) by adding the constraint
$$\ar a=b$$
is equivalent to the conjunction of the completion
sentences~(\ref{a2comp})
with $a\neq b$.  This example illustrates the role of a property more general
than the logical equivalence between $\sm[F]$ and the completion of~$F$:
it may be useful to know when the equivalence between these two formulas is
entailed by a certain set of assumptions.  This information may be relevant if
we are interested in a logic program obtained from~$F$ by adding constraints.

The result of applying~$\sm$ to the program
\beq
\ba {rl}
p(a)\!\!& \ar p(b),\\
q(c)\!\!& \ar q(d),\\
   \!\!& \ar a=b,\\
   \!\!& \ar c=d
\ea
\eeq{a3}
is equivalent to the conjunction of the formulas
\beq
\ba c
\forall x(p(x)\lrar x=a\land p(b)),\\
\forall x(q(x)\lrar x=c\land q(d)),\\
a\neq b,\\
c\neq d.
\ea
\eeq{a3comp}
This claim cannot be justified, however, by a reference to Corollary~5 from
\cite{lee11b}.  The program in this example is atomic-tight, but it does not
contain constraints  corresponding to some of the unique name axioms, for
instance $a\neq c$.  We will show how our claim follows from the main theorem
stated below.

We will discuss also an example illustrating limitations of earlier work that
is related to describing dynamic domains in answer set programming (ASP).  The
program in that example is not atomic-tight because of rules expressing the
commonsense law of inertia.  We will show nevertheless that the process of
completion can be used to characterize its stable models by a first-order
formula.

The class of tight programs is defined in~\cite{fer09} in terms of
predicate dependency graphs; that definition is reproduced in
Section~\ref{sec:rev2}
below.  The definition of an atomic-tight program in \cite{lee11b} refers
to more informative ``first-order dependency graphs.''  Our approach is based
on an alternative solution to the problem of making predicate dependency graphs
more informative, ``rule dependency graphs.''

After reviewing some background material in Sections~\ref{sec:rev1}
and~\ref{sec:rev2}, we define rule dependency graphs in Section~\ref{sec:rdg},
state the main theorem and give examples of its use in Sections~\ref{sec:main}
and~\ref{sec:larger}, and outline its proof in
Sections~\ref{sec:inf} and~\ref{sec:proof}.

\section{Review: Operator SM, Lloyd-Topor Programs, and Completion}
\label{sec:rev1}

In this paper, a {\em formula} is a first-order formula that may
contain the
propositional connectives~$\bot$ (logical falsity), $\land$, $\lor$,
and~$\rar$, and the quantifiers~$\forall$,~$\exists$.  We treat $\neg
F$ as
an abbreviation for $F\rar\bot$;~$\top$ stands for $\bot\rar\bot$;
$F\lrar G$ stands for $(F\rar G)\land(G\rar F)$.

For any first-order sentence $F$ and any tuple~$\bf p$ of distinct
predicate constants (other than equality)
$\sm_{\bf p}[F]$ is the conjunction of~$F$ with a second-order "stability
condition"; see \cite[Section~2]{fer09} for details.  The members of~$\bp$
are called {\em intensional}, and the other predicate constants are
{\em extensional}.
We will drop the subscript in the symbol $\sm_{\bf p}$ when $\bf p$ is the
list of all predicate symbols occurring in~$F$.
For any sentence~$F$, a {\em $\bf p$-stable} (or simply {\em
stable\/}) {\em
model} of~$F$ is an interpretation of the underlying signature that
satisfies $\sm_{\bf p}[F]$.

A {\em Lloyd-Topor program} is a finite set of rules of the form
\beq
p(\bt)\ar G,
\eeq{rule}
where $\bt$ is a tuple of terms, and $G$ is a formula.  We will
identify a
program with the sentence obtained by conjoining the formulas
$$
\widetilde\forall(G\rar p(\bt))
$$
for all its rules~(\ref{rule}). ($\widetilde\forall F$ stands for the
universal closure of~$F$.)

Let $\Pi$ be a Lloyd-Topor program, and $p$ a predicate constant
(other than equality).
Let
\beq
p(\bt^i)\ar G^i\qquad (i=1,2,\dots)
\eeq{rule2}
be all rules of $\Pi$ that contain~$p$ in the head.  The {\em definition
of~$p$ in~$\Pi$} is the rule
\beq
p(\bx)\ar \bigvee_{i}\exists \by^i (\bx=\bt^i\land G^i),
\eeq{def}
where $\bx$ is a list of distinct variables not appearing in any of
the
rules~(\ref{rule2}), and $\by^i$ is the list of free variables
of~(\ref{rule2}).\footnote{By $\bx=\bt^i$ we denote the conjunction
of the
equalities between members of the tuple $\bx$ and the corresponding
members
of the tuple $\bt^i$.}   The {\em
completed
definition of $p$ in $\Pi$} is the formula
\beq
\forall\bx\left(p(\bx)\lrar \bigvee_{i}\exists \by^i (\bx=\bt^i\land
G^i)\right).
\eeq{comp}
 For instance, the completed definitions of~$p$
and~$q$
in program~(\ref{a1}) are the formulas
$$
\ba l
\forall x_1(p(x_1)\lrar x_1=a \lor \exists x(x_1=x\land q(x))),\\
\forall x_1(q(x_1)\lrar x_1=b),
\ea
$$
which can be equivalently rewritten as~(\ref{a1comp}).

By $\comp[\Pi]$ we denote the conjunction of the completed definitions
of
all predicate constants~$p$ in~$\Pi$.  This sentence is similar to the
completion of~$\Pi$ in the sense of~\cite[Section~2]{llo84a}, except
that it
does not include Clark equality axioms.

\section{Review: Tight Programs}\label{sec:rev2}

We will review now the definition of tightness from \cite[Section~7.3]{fer09}.
In application to a Lloyd-Topor program~$\Pi$, when
all predicate constants occurring in~$\Pi$ are treated as intensional, that
definition can be stated as follows.

An occurrence of an expression in a first-order formula is~{\em negated} if
it belongs to a subformula of the form $\neg F$ (that is, $F\rar\bot$), and
{\em nonnegated} otherwise.  The {\em predicate dependency graph
of~$\Pi$} is the directed graph that has
\begin{itemize}
\item
all predicate constants occurring in~$\Pi$ as its vertices, and
\item
an edge from~$p$ to~$q$ whenever~$\Pi$ contains a rule~(\ref{rule}) with~$p$
in the head such that its body~$G$ has a positive\footnote{Recall that an
occurrence of an expression in a first-order formula is called~{\em positive}
if the number of implications containing that occurrence in the antecedent is
even.} nonnegated occurrence of~$q$.
\end{itemize}
We say that~$\Pi$ is {\em tight} if the predicate dependency graph of~$\Pi$
is acyclic.

For example, the predicate dependency graph of program~(\ref{a1}) has a
single edge, from~$p$ to~$q$.  The predicate dependency graph of
program~(\ref{a2}) has two edges, from~$p$ to~$q$ and from~$q$ to~$p$.
The predicate dependency graph of the
program
\beq
\ba l
p(a,b)\\
q(x,y)\ar p(y,x)\land\neg p(x,y)
\ea
\eeq{ex1}
has a single edge, from~$q$ to~$p$ (because one of the occurrences of~$p$ in
the body of the second rule is nonnegated).  The predicate dependency graph
of the program
\beq
\ba l
p(x)\ar q(x),\\
q(x)\ar r(x),\\
r(x)\ar s(x)
\ea
\eeq{ex2}
has 3 edges:
$$
p
\;\longrightarrow\;
q
\;\longrightarrow\;
r
\;\longrightarrow\;
s.$$
Programs~(\ref{a1}),~(\ref{ex1}) and~(\ref{ex2}) are tight; program~(\ref{a2})
is not.

\begin{proposition} \label{prop1}
$\;$If a Lloyd-Topor program~$\Pi$ is tight then $\sm[\Pi]$ is equivalent to
$\comp[\Pi]$.
\end{proposition}

This is an easy corollary to the theorem from \cite{fer09}
mentioned in the introduction.  Indeed, consider the set~$\Pi'$ of the
definitions~(\ref{def}) of all predicate constants~$p$ in~$\Pi$.  It can be
viewed as a
formula in Clark normal form in the sense of~\cite[Section~6.1]{fer09}.
It is tight, because it has the same predicate dependency graph as~$\Pi$.
By Theorem~11 from~\cite{fer09}, $\sm[\Pi']$ is equivalent to the completion
of~$\Pi'$ in the sense of~\cite[Section~6.1]{fer09}, which is identical to
$\comp[\Pi]$.  It remains to observe that~$\Pi$ is intuitionistically
equivalent to~$\Pi'$, so that~$\sm[\Pi]$ is equivalent to~$\sm[\Pi']$
\cite[Section~5.1]{fer09}.

\section{Rule Dependency Graph}\label{sec:rdg}

We are interested in conditions on a Lloyd-Topor program~$\Pi$ ensuring
that the equivalence
$$
\sm[\Pi]\lrar\comp[\Pi]
$$
is entailed by a given set of assumptions~$\Gamma$.  Proposition~\ref{prop1}
gives a solution for the special case when~$\Gamma$ is empty.  The following
definition will help us answer the more general question.

The {\em rule dependency graph} of a Lloyd-Topor program~$\Pi$ is the
directed graph that has
\begin{itemize}
\item
rules of~$\Pi$, with variables (both free and bound) renamed arbitrarily, as
its vertices, and
\item an edge from a rule $p({\bf t}) \ar G$ to a
rule $p'({\bf t}') \ar G'$, labeled by an atomic formula $p'({\bf s})$, if
$p'({\bf s})$ has a  positive nonnegated occurrence in $G$.
\end{itemize}

Unlike the predicate dependency graph, the rule dependency graph of a program
is usually infinite.
For example, the rule dependency graph of program~(\ref{ex1}) has the vertices
$p(a,b)$ and
\beq
q(x_1,y_1)\ar p(y_1,x_1)\land\neg p(x_1,y_1)
\eeq{v}
for arbitrary pairs of distinct variables $x_1,y_1$.  It has an edge
from each vertex~(\ref{v}) to $p(a,b)$, labeled $p(y_1,x_1)$.  The rule
dependency graph of program~(\ref{ex2}) has edges of two kinds:
\begin{itemize}
\item from $p(x_1)\ar q(x_1)$ to $q(x_2)\ar r(x_2)$, labeled~$q(x_1)$, and
\item from $q(x_1)\ar r(x_1)$ to $r(x_2)\ar s(x_2)$, labeled~$r(x_1)$
\end{itemize}
for arbitrary variables~$x_1$,~$x_2$.

The rule dependency graph of a program is ``dual'' to its predicate
dependency graph, in the following sense.  The vertices of the predicate
dependency graph are predicate symbols, and the presence of an edge from~$p$
to~$q$ is determined by the existence of a rule that contains certain
occurrences of~$p$ and~$q$.  The vertices of the rule dependency graph are
rules, and the presence of an edge from~$R_1$ to~$R_2$ is determined by the
existence of a predicate symbol with certain occurrences in~$R_1$ and~$R_2$.

There is a simple characterization of tightness in terms of rule dependency
graphs:

\begin{proposition}
A Lloyd-Topor program~$\Pi$ is tight iff there exists~$n$
such that the rule dependency graph of~$\Pi$ has no paths of length~$n$.
\end{proposition}

\noindent{\bf Proof.}  Assume that~$\Pi$ is tight, and let~$n$ be
the number of predicate symbols occurring in~$\Pi$.  Then the rule dependency
graph of~$\Pi$ has no paths of length~$n+1$.  Indeed, assume that such a
path exists:
$$R_0
\;{\,p_1(\dots) \over {\;}}\!\!\!\!\!\!\!\!\rightarrow
R_1
\;{\,p_2(\dots) \over {\;}}\!\!\!\!\!\!\!\!\rightarrow
R_2
\;{\,p_3(\dots) \over {\;}}\!\!\!\!\!\!\!\!\rightarrow
\cdots
\;{\,p_{n+1}(\dots) \over {\;}}\!\!\!\!\!\!\!\!\rightarrow
R_{n+1}.
$$
Each of the rules~$R_i$ ($1\leq i\leq n$) contains~$p_i$ in the head and a
positive nonnegated occurrence of~$p_{i+1}$ in the body.  Consequently the
predicate dependency graph of~$\Pi$ has an edge from~$p_i$ to~$p_{i+1}$, so
that $p_1,\dots,p_{n+1}$ is a path in that graph; contradiction.
Now assume that~$\Pi$ is not tight.  Then there is an infinite path
$p_1,p_2,\dots$ in the predicate dependency graph of~$\Pi$.  Let~$R_i$ be a
rule of~$\Pi$ that has~$p_i$ in the head and a positive nonnegated occurrence
of~$p_{i+1}$ in the body.  Then the rule dependency graph of~$\Pi$ has an
infinite path of the form
$$
R_1
\;{\,p_2(\dots) \over {\;}}\!\!\!\!\!\!\!\!\rightarrow
R_2
\;{\,p_3(\dots) \over {\;}}\!\!\!\!\!\!\!\!\rightarrow
\cdots\;.
$$

The main theorem, stated in the next section, refers to finite paths in
the rule dependency graph of a program~$\Pi$ that satisfy an additional
condition: the rules at their vertices have no common variables (neither free
nor bound).  Such paths will be called {\em chains}.

\begin{corollary} \label{cor1}
A Lloyd-Topor program~$\Pi$ is tight iff there exists~$n$
such that~$\Pi$ has no chains of length~$n$.
\end{corollary}

Indeed, any finite path in the rule dependency graph of~$\Pi$ can be
converted into a chain of the same length by renaming variables.

\section{Main Theorem} \label{sec:main}

Let~$C$ be a chain
\beq
\ba c
p_0({\bf t}^0) \ar \i{Body}_0\\
\hbox{\Large $\downarrow$} p_1({\bf s}^1)\\
p_1({\bf t}^1) \ar \i{Body}_1\\
\hbox{\Large $\downarrow$} p_2({\bf s}^2)\\
%\cdots\\
.\;.\;.\;.\;.\;.\;.\;.\;.\;.\;.\\
\hbox{\Large $\downarrow$} p_n({\bf s}^n)\\
p_n({\bf t}^n) \ar \i{Body}_n
\ea
\eeq{chain}
in a Lloyd-Topor program~$\Pi$.
The corresponding {\em chain formula} $F_C$ is the conjunction
$$\bigwedge_{i=1}^n{\bf s}^i={\bf t}^i\,\land\bigwedge_{i=0}^n\i{Body}_i.$$
For instance, if~$C$ is the chain
$$\ba c\
q(x_1,y_1)\ar p(y_1,x_1)\land \neg p(x_1,y_1)\\
\hbox{\Large $\downarrow$} p(y_1,x_1)\\
\!\!\!\!\!\!\!\!\!\!\!\!\!\!\!\!\!\!\!\!\!\!
p(a,b)
\ea
$$
in program~(\ref{ex1}) then~$F_C$ is
$$y_1=a\land x_1=b \land p(y_1,x_1) \land \neg p(x_1,y_1).$$

Let~$\Gamma$ be a set of sentences.  About a Lloyd-Topor program~$\Pi$ we
will say that it is {\em tight relative to~$\Gamma$}, or {\em $\Gamma$-tight},
if there exists a positive
integer~$n$ such that, for every chain~$C$ in~$\Pi$ of length~$n$,
$$
\Gamma, \hbox{\rm Comp}[\Pi] \models \uc\neg F_C.
$$

\medskip\noindent{\bf Main Theorem.}
{\em
If a Lloyd-Topor program~$\Pi$ is $\Gamma$-tight then
$$
\Gamma\models\hbox{\rm SM}[\Pi]\lrar\hbox{\rm Comp}[\Pi].
$$
}

Corollary~1 shows that every tight program is trivially $\Gamma$-tight even
when~$\Gamma$ is empty.  Consequently the main theorem can be viewed as a
generalization of Proposition~\ref{prop1}.

Tightness in the sense of Section~\ref{sec:rev2} is a syntactic condition
that is easy to verify; $\Gamma$-tightness is not.  Nevertheless, the main
theorem is useful because it may allow us to reduce the problem of
characterizing the stable models of a program by a first-order formula
to verifying an entailment in first-order logic.

Here are some examples.  In each case, to
verify $\Gamma$-tightness we take $n=1$.  We will check the entailment in the
definition of $\Gamma$-tightness by deriving a contradiction from (some subset
of) the assumptions $\Gamma$, $\hbox{\rm Comp}[\Pi]$, and~$F_C$.

\medskip\noindent{\bf Example 1.} The one-rule program
$$
p(a)\ar p(x)\land x\neq a
$$
is tight relative to~$\emptyset$.  Indeed, any chain of length~1 has the form
$$
\ba c
p(a)\ar p(x_1)\land x_1\neq a\\
\hbox{\Large $\downarrow$} p(x_1)\\
p(a)\ar p(x_2)\land x_2\neq a.
\ea
$$
The corresponding chain formula
$$x_1=a\land p(x_1)\land x_1\neq a\land p(x_2)\land x_2\neq a.$$
is contradictory.

Thus the stable models of this program are described by its completion,
even though the program is not tight (and not even atomic-tight).

\medskip\noindent{\bf Example 2.} Let~$\Pi$ be the program consisting of
the first 2 rules of~(\ref{a3}):
$$
\ba c
p(a)\ar p(b),\\
q(c)\ar q(d).
\ea
$$
To justify the claim about~(\ref{a3}) made in the introduction, we will
check that~$\Pi$ is tight relative to $\{a\neq b,c\neq d\}$.
There are two chains of length~1:
$$\ba c\
p(a)\ar p(b)\\
\hbox{\Large $\downarrow$} p(b)\\
p(a)\ar p(b)
\ea
$$
and
$$\ba c\
q(c)\ar q(d)\\
\hbox{\Large $\downarrow$} q(d)\\
q(c)\ar q(d).
\ea
$$
The corresponding chain formulas are
$$b=a\land p(b)\land p(b)$$
and
$$d=c\land q(d)\land q(d).$$
Each of them contradicts~$\Gamma$.

\medskip\noindent{\bf Example 3.} Let us check that program~(\ref{a2}) is
tight relative to $\{a\neq b\}$.  Its chains of length~1 are
$$
\ba c\
p(x_1)\ar q(x_1)\\
\hbox{\Large $\downarrow$} q(x_1)\\
q(a)\ar p(b)
\ea
$$
and
$$
\ba c\
q(a)\ar p(b)\\
\hbox{\Large $\downarrow$} q(b)\\
p(x_1)\ar q(x_1)
\ea
$$
for an arbitrary variable~$x_1$.  The corresponding chain formulas include
the conjunctive term~$p(b)$.  Using the completion~(\ref{a2comp}) of the
program, we derive $b=a$, which contradicts~$\Gamma$.

\section{A Larger Example}\label{sec:larger}

Programs found in actual applications of ASP usually involve constructs that
are not allowed in Lloyd-Topor programs, such as choice rules and
constraints.  Choice rules have the form
$$\{p(\bt)\}\ar G.$$
We view this expression as shorthand for the sentence
$$
\widetilde\forall(G\rar p(\bt)\lor\neg p(\bt)).
$$
A constraint $\ar G$ is shorthand for the sentence $\widetilde\forall\neg G$.
Such sentences do not correspond to any rules in the sense of
Section~\ref{sec:rev1}.

Nevertheless, the main theorem stated above can sometimes
help us characterize the stable models of a ``realistic'' program by a
first-order formula.  In this section we discuss an example of this kind.

The logic program~$M$ described below encodes commonsense knowledge about
moving objects from one location to another.  Its signature consists of
\begin{itemize}
\item  the object constants $\widehat{0}, \ldots, \widehat{k}$, where~$k$ is
a fixed nonnegative integer;
\item the unary predicate constants $\i{object}$, $\i{place}$, and \i{step};
they correspond to the three types of individuals under consideration;
\item the binary predicate constant $\i{next}$; it describes the temporal
order of steps;
\item the ternary predicate constants $\i{at}$ and $\i{move}$; they represent
the fluents and actions that we are interested in.
\end{itemize}
The predicate constants \i{step}, \i{next}, and \i{at} are intensional;
the other three are not. (The fact that some predicates are extensional
is the first sign that~$M$ is not a Lloyd-Topor program.)  The program
consists of the following rules:
\begin{enumerate}
\item[(i)] the facts
$$
\ba{l}
\i{step}(\widehat{0}),\; \i{step}(\widehat{1}),\;\ldots\;\i{step}(\widehat{k});\\
\i{next}(\widehat{0},\widehat{1}),\;
\i{next}(\widehat{1},\widehat{2}),\;\ldots,\;
\i{next}(\widehat{k\!-\!1},\widehat{k});
\ea
$$
\item[(ii)] the unique name constraints
$$\ar \widehat{i}=\widehat{j}\qquad(1\le i< j\le k);$$
\item[(iii)] the constraints describing the arguments of \i{at} and \i{move}:
$$
\ar \i{at}(x,y,z)\land\neg(\i{object}(x)\land \i{place}(y)\land \i{step}(z))
$$
and
$$
\ar \i{move}(x,y,z)\land\neg(\i{object}(x)\land \i{place}(y)\land \i{step}(z));
$$
\item[(iv)] the uniqueness of location constraint
$$\ar\i{at}(x,y_1,z)\land\i{at}(x,y_2,z)\land y_1\neq y_2;$$
\item[(v)] the existence of location constraint
$$\ar \i{object}(x)\land\i{step}(z)\land \neg\exists y~\i{at}(x,y,z);$$
\item[(vi)] the rule expressing the effect of moving an object:
$$\i{at}(x,y,u) \ar \i{move}(x,y,z)\land\i{next}(z,u);$$
\item[(vii)] the choice rule expressing that initially an object can be
anywhere:
$$\{\i{at}(x,y,0)\} \ar \i{object}(x)\land\i{place}(y);$$
\item[(viii)] the choice rule expressing the commonsense law of
inertia:\footnote{This representation of inertia follows the example of
\cite[Figure~1]{lee12}.}
$$\{\i{at}(x,y,u)\} \ar \i{at}(x,y,z)\land \i{next}(z,u).$$
\end{enumerate}

Program~$M$ is not atomic-tight, so that methods of \cite{lee11b} are not
directly applicable to it.  Nevertheless, we can describe the stable models
of this program without the use of second-order quantifiers.  In the
statement of the proposition below, $\bp$ stands for the list of
intensional predicates $\i{step}$, $\i{next}$ and $\i{at}$, and~$H$ is
the conjunction of the universal closures of the formulas
$$\ba{l}
\widehat{i}\neq \widehat{j} \qquad(1\le i<j\le k),\\
\i{at}(x,y,z)\rar \i{object}(x)\land \i{place}(y)\land \i{step}(z),\\
\i{move}(x,y,z)\rar \i{object}(x)\land \i{place}(y)\land \i{step}(z),\\
\i{at}(x,y_1,z)\land\i{at}(x,y_2,z)\rar y_1=y_2,\\
\i{object}(x)\land\i{step}(z)\rar \exists y~\i{at}(x,y,z).
\ea$$

\begin{proposition}\label{prop3}
$\sm_{\scriptsize \bp}[M]$ is equivalent to the
conjunction of $H$ with the universal closures of the formulas
\beq
\i{step}(z)\lrar \bigvee_{i=0}^{k}z=\widehat{i},
\eeq{f3step}
\beq
\i{next}(z,u)\lrar
\bigvee_{i=0}^{k-1}(z=\widehat{i}\land u=\widehat{i\!+\!1}),
\eeq{f3next}
\beq
\ba l
\i{at}(x,y,\widehat{i\!+\!1})\lrar
(\i{move}(x,y,\widehat{i})
\lor (\i{at}(x,y,\widehat i)\land\neg\exists w~move(x,w,\widehat{i})))\\
\qquad\qquad\qquad\qquad\qquad\qquad\qquad\qquad\qquad\qquad\qquad
(i=0,\dots,k-1).
\ea
\eeq{f3at}
\end{proposition}

Recall that the effect of adding a constraint to a logic program is to
eliminate its stable models that violate that constraint
\cite[Theorem~3]{fer09}.  An interpretation satisfies~$H$ iff it does not
violate any of the constraints~(ii)--(v).  So the statement of
Proposition~\ref{prop3} can be summarized as follows:
the contribution of rules~(i) and (vi)--(viii), under the
stable model semantics, amounts to providing explicit definitions for \i{step}
and \i{next}, and ``successor state formulas'' for \i{at}.

The proof of Proposition~\ref{prop3} refers to the Lloyd-Topor program~$\Pi$
consisting of rules~(i), (vi),
\begin{enumerate}
\item[(vii$'$)]
$\i{at}(x,y,0) \ar \i{object}(x)\land\i{place}(y)\land\neg\neg \i{at}(x,y,0)$,
\item[(viii$'$)]
$\i{at}(x,y,u) \ar \i{at}(x,y,z)\land \i{next}(z,u)
\land\neg\neg\i{at}(x,y,t_2)$,
\end{enumerate}
and
\beq
\ba {rl}
\i{object}(x)\!\!&\ar\neg\neg\i{object}(x),\\
\i{place}(y)\!\!&\ar\neg\neg\i{place}(y),\\
\i{move}(x,y,z)\!\!&\ar\neg\neg\i{move}(x,y,z).
\ea
\eeq{ch}
It is easy to see that $\sm_{\bf p}[M]$ is equivalent to
$\sm[\Pi]\land H$.  Indeed, consider the program~$M'$ obtained from~$M$
by adding rules~(\ref{ch}).  These rules are strongly equivalent to the
choice rules
$$\{\i{object}(x)\},\;\{\i{place}(y)\},\;\{\i{move}(x,y,z)\}.$$
Consequently $\sm_{\bf p}[M]$ is equivalent to $\sm[M']$
\cite[Theorem~2]{fer09}.  It remains to notice that (vii) is strongly
equivalent to (vii$'$), and
(viii) is strongly equivalent to (viii$'$).

Furthermore---and this is the
key step in the proof of Proposition~\ref{prop3}---the second-order formula
$\sm[\Pi]\land H$ is equivalent to the first-order formula
$\comp[\Pi]\land H$, in view of our main theorem and the following fact:

\begin{lemma}\label{lemmatoprop3}
Program~$\Pi$ is $H$-tight.
\end{lemma}

To derive Proposition~\ref{prop3} from the lemma, we only need to
observe that~(\ref{f3step})
and~(\ref{f3next}) are the completed definitions of \i{step} and \i{next}
in~$\Pi$, and that the completed definition of \i{at} can be transformed
into~(\ref{f3at}) under assumptions~(\ref{f3step}),~(\ref{f3next}), and~$H$.

\medskip\noindent{\bf Proof of Lemma~\ref{lemmatoprop3}.}
Consider a chain in~$\Pi$ of length $k+2$:
\beq
R_0
\;{\,p_1(\dots) \over {\;}}\!\!\!\!\!\!\!\!\rightarrow
R_1
\;{\,p_2(\dots) \over {\;}}\!\!\!\!\!\!\!\!\rightarrow
\cdots
\;{\,p_{k+1}(\dots) \over {\;}}\!\!\!\!\!\!\!\!\rightarrow
R_{k+1}
\;{\,p_{k+2}(\dots) \over {\;}}\!\!\!\!\!\!\!\!\rightarrow
R_{k+2}.
\eeq{lc}
Each~$R_i$ is obtained from one of the
rules~(i),~(vi),~(vii$'$),~(viii$'$),~(\ref{ch}) by renaming variables.
Each~$p_i$ occurs in the head of~$R_i$ and has a
positive nonnegated occurrence in~$R_{i-1}$.  Since there are no nonnegated
predicate symbols in the bodies of rules~(i) and~(\ref{ch}), we conclude that
$R_0,\dots,R_{k+1}$ are obtained from other rules of~$\Pi$, that is, from
(vi),~(vii$'$), and~(viii$'$).  Since the predicate constant in the head of
each of these three rules is~\i{at}, each of~$p_1,\dots,p_{k+1}$ is the
symbol~\i{at}.  Since there are no nonnegated occurrences of \i{at} in
the bodies of~(vi) and~(vii$'$), we conclude that $R_0,\dots,R_k$ are
obtained by renaming variables in~(viii$'$).  This means that chain~(\ref{ch})
has the form
$$\ba c\
\i{at}(x_0,y_0,u_0)\ar \i{at}(x_0,y_0,z_0)\land \i{next}(z_0,u_0)
\land\neg\neg\i{at}(x_0,y_0,u_0)\\
\qquad\qquad\hbox{\Large $\downarrow$} \i{at}(x_0,y_0,z_0)\\
\i{at}(x_1,y_1,u_1)\ar \i{at}(x_1,y_1,z_1)\land \i{next}(z_1,u_1)
\land\neg\neg\i{at}(x_1,y_1,u_1)\\
\qquad\qquad\hbox{\Large $\downarrow$} \i{at}(x_1,y_1,z_1)\\
\!\!\!\!\!\!\!\!\ldots\\
\qquad\qquad\qquad\quad\hbox{\Large $\downarrow$} \i{at}(x_{k-1},y_{k-1},z_{k-1})\\
\i{at}(x_k,y_k,u_k)\ar \i{at}(x_k,y_k,z_k)\land \i{next}(z_k,u_k)
\land\neg\neg \i{at}(x_k,y_k,u_k)\\
\qquad\qquad\hbox{\Large $\downarrow$} \i{at}(x_k,y_k,z_k)\\
R_{k+1}\\
\hbox{\Large $\downarrow$} \cdots\\
R_{k+2}.
\ea
$$
The corresponding chain formula contains the conjunctive terms
$$
z_0=u_1,z_1=u_2,\dots,z_{k-1}=u_k
$$
and
$$
\i{next}(z_0,u_0),\i{next}(z_1,u_1),\ldots,\i{next}(z_k,u_k).
$$
From these formulas we derive
\beq
\i{next}(u_1,u_0),\i{next}(u_2,u_1),\ldots,\i{next}(u_{k+1},u_k),
\eeq{next2}
where $u_{k+1}$ stands for~$z_k$.  Using the completed definition of
\i{next}, we conclude:
$$u_i=\widehat 0 \lor \cdots\lor u_i=\widehat k\qquad (0\leq i\leq k+1).$$
Consider the case when
$$u_i=\widehat {j_i}\qquad (0\leq i\leq k+1)$$
for some numbers $j_0,\dots,j_{k+1}\in\{0,\dots,k\}$.
There exists at least one subscript~$i$ such that $j_i\neq j_{i+1}+1$,
because otherwise we would have
$$j_0=j_1+1=j_2+2=\cdots=j_{k+1}+k+1,$$
which is impossible because $j_0,j_{k+1}\in\{0,\dots,k\}$.  By the choice
of~$i$, from the completed definition of \i{next} and the unique name assumption
(included in~$H$) we can derive
$\neg\i{next}(\widehat {j_{i+1}},\widehat {j_i})$.
Consequently $\neg\i{next}(u_{i+1},u_i)$, which
contradicts~(\ref{next2}).

\section{Review: Stable Models of Infinitary Formulas}\label{sec:inf}

Our proof of the main theorem employs the method proposed (for a different
purpose) by Miroslaw Truszczynski \cite{tru12}, and in this section we
review some of the definitions and results of that paper.  The stable model
semantics of propositional formulas due to Paolo Ferraris \cite{fer05}
is extended there to formulas with infinitely long conjunctions and
disjunctions, and that generalization is related to the operator~\sm.

Let $\mathcal A$ be a set of propositional atoms.  The sets
$\mathcal{F}_0,\mathcal{F}_1,\dots$ are defined as follows:
\begin{itemize}
\item $\mathcal{F}_0=\mathcal A\cup\{\bot\}$;
\item $\mathcal{F}_{i+1}$ consists of expressions
$\mathcal{H}^\lor$ and $\mathcal{H}^\land$, for all subsets
$\mathcal{H}$ of $\mathcal{F}_0\cup\ldots\cup\mathcal{F}_i$, and of
expressions $F\rar G$, where $F,G\in\mathcal{F}_0\cup\ldots\cup\mathcal{F}_i$.
\end{itemize}
An {\sl infinitary formula} (over $\mathcal A$) is an element of
$\bigcup^{\infty}_{i=0}\mathcal{F}_i$.

A {\em (propositional) interpretation} is a subset of $\mathcal A$.  The
satisfaction relation between an interpretation and an infinitary formula is
defined in a natural way.  The definition of the reduct~$F^I$ of a formula~$F$
relative to an interpretation~$I$ proposed in \cite{fer05} is extended to
infinitary formulas as follows:
\begin{itemize}
\item $\bot^I=\bot$.
\item For $A\in \mathcal A$, $A^I=\bot$ if $I\not\models A$; otherwise
$A^I=A$.
\item $(\mathcal{H}^\land)^I=\bot$ if $I\not\models
\mathcal{H}^\land$; otherwise
$(\mathcal{H}^\land)^I=\{G^I|G\in\mathcal{H}\}^\land$.
\item $(\mathcal{H}^\lor)^I=\bot$ if $I\not\models
\mathcal{H}^\lor$; otherwise
$(\mathcal{H}^\lor)^I=\{G^I|G\in\mathcal{H}\}^\lor$.
\item $(G\rar H)^I=\bot$ if $I\not\models G\rar H$; otherwise
$(G\rar H)^I=G^I\rar H^I$.
\end{itemize}
An interpretation $I$ is a {\em stable model} of an infinitary formula~$F$ if
$I$ is a minimal model of~$F^I$.  An interpretation~$I$ satisfies~$F^I$ iff it
satisfies~$F$ \cite[Proposition~1]{tru12}, so that stable models of~$F$ are
models of~$F$.

Infinitary formulas are used to encode first-order sentences as follows.
For any interpretation~$I$ in the sense of first-order logic, let~$\mathcal A$
be the
set of ground atoms formed from the predicate constants of the underlying
signature and the ``names'' $\xi^*$ of elements~$\xi$ of the universe $|I|$
of~$I$---new objects constants that are in a 1--1 correspondence with elements
of~$|I|$.  By $I^r$ we denote the set of atoms from~$\mathcal A$ that are
satisfied
by~$I$.  In the definition below, $t^I$ stands for the value
assigned to the ground term~$t$ by the interpretation~$I$.  The {\em grounding}
of a first-order sentence~$F$ relative to~$I$ (symbolically, $\gr_I(F)$) is the
infinitary formula over~$A$ constructed as follows:
\begin{itemize}
\item $\gr_I(\bot)=\bot$.
\item $\gr_I(p(t_1,\ldots, t_k))=p((t^I_1)^*,\ldots, (t^I_k)^*)$.
\item $\gr_I(t_1=t_2)=\top$, if $t_1^I=t_2^I$, and $\bot$ otherwise.
\item If $F=G\lor H$, $\gr_I(F)=\gr_I(G)\lor \gr_I(H)$ (the case of $\land$
is analogous).
\item If $F=G\rar H$, $\gr_I(F)=\gr_I(G)\rar \gr_I(H)$.
\item If $F=\exists x G(x)$, $\gr_I(F)=\{\gr_I(G(u^*))|u\in |I|\}^\lor$.
\item If $F=\forall x G(x)$, $\gr_I(F)=\{\gr_I(G(u^*))|u\in |I|\}^\land$.
\end{itemize}
It is easy to check that $\gr_I$ is a faithful translation in the following
sense: $I$ satisfies a first-order sentence~$F$ iff~$I^r$ satisfies
$\gr_I(F)$.

This transformation is also faithful in the sense of the stable model semantics:
$I$ satisfies $\sm[F]$ iff~$I^r$ is a stable model of $\gr_I(F)$
\cite[Theorem~5]{tru12}.  This is why
infinitary formulas can be used for proving properties of the operator~$\sm$.

\section{Proof Outline}\label{sec:proof}

In the statement of the main theorem, the implication left-to-right
$$
\hbox{\rm SM}[\Pi]\rar\hbox{\rm Comp}[\Pi]
$$
is logically valid for any Lloyd-Topor program~$\Pi$.  This fact
follows from \cite[Theorem~11]{fer09} by the argument used in the
proof of Proposition~\ref{prop1} above.  In this section we outline the
proof in the other direction:

\medskip
\begin{center}{\em
If a Lloyd-Topor program~$\Pi$ is $\Gamma$-tight,\\
and an interpretation~$I$ satisfies both $\Gamma$ and $\comp[\Pi]$,\\
then~$I$ satisfies $\sm[\Pi]$.
}\end{center}

\medskip
This assertion follows from three lemmas.  The first of them expresses a
Fages-style
property of infinitary formulas similar to Theorem 1 from \cite{erd03}.  It
deals with {\em infinitary programs\/}---conjunctions of (possibly
infinitely many) implications $G\rar A$ with \hbox{$A\in\mathcal A$}.
Such an implication will be called an {\em (infinitary) rule} with the
{\em head}~$A$ and {\em body}~$G$, and we will write it as $A\ar G$.
For instance, if~$\Pi$ is a Lloyd-Topor program then, for any
interpretation~$I$, $\gr_I(\Pi)$ is an infinitary program. We say that
an interpretation $I$ is {\em supported} by an infinitary program~$\Pi$ if
each atom $A\in I$ is the head of a rule~$A\ar G$ of~$\Pi$ such that
$I\models G$.  The lemma shows that under some condition the
stable models of an infinitary program~$\Pi$ can be characterized as the
interpretations that satisfy~$\Pi$ and are supported by~$\Pi$.

The condition refers to the set of {\em positive nonnegated atoms} of an
infinitary formula.  This set,
denoted by~$\pnn(F)$, and the set of {\em negative nonnegated atoms} of~$F$,
denoted by~$\nnn(F)$, are defined recursively, as follows:
\begin{itemize}
\item $\pnn(\bot)=\emptyset$.
\item For $A\in\mathcal A$, $\pnn(A)=\{A\}$.
\item $\pnn(\mathcal{H}^\land)=\pnn(\mathcal{H}^\lor)
       =\bigcup_{H\in\mathcal{H}}\pnn(H)$.
\item $\pnn(G\rar H)=\left\{\ba{ll}\emptyset & \mbox{if}\ H=\bot, \\
\nnn(G)\cup\pnn(H) & \hbox{otherwise.}\ea\right.$
\end{itemize}
\begin{itemize}
\item $\nnn(\bot)=\emptyset$,
\item For $A\in\mathcal A$, $\nnn(A)=\emptyset$.
\item $\nnn(\mathcal{H}^\land)=\nnn(\mathcal{H}^\lor)
       =\bigcup_{H\in\mathcal{H}}\nnn(H)$.
\item $\nnn(G\rar H)=\left\{\ba{ll}\emptyset & \hbox{if}\ H=\bot, \\
\pnn(G)\cup\nnn(H) & \hbox{otherwise.}\ea\right.$
\end{itemize}

Let~$\Pi$ be an infinitary program, and~$I$ a propositional interpretation.
About atoms $A,A'\in I$ we say that $A'$ is a {\em parent of $A$
relative to $\Pi$ and $I$} if~$\Pi$ has a rule $A\ar G$ with the head~$A$
such that $I\models G$ and $A'$ is a positive nonnegated
atom of~$G$.  We say that~$\Pi$
is {\em tight on $I$} if there is no infinite sequence $A_0, A_1, \ldots $ of
elements of~$I$ such that for every~$i$, $A_{i+1}$ is a parent of $A_i$
relative to~$F$ and~$I$.

\begin{lemma}\label{lem:gen}
For any model~$I$ of an infinitary program~$\Pi$ such that~$\Pi$ is tight
on~$I$, $I$ is stable iff~$I$ is supported by~$\Pi$.
\end{lemma}

The next lemma relates the $\Gamma$-tightness condition from the statement
of the main theorem to tightness on an interpretation defined above.

\begin{lemma}\label{lem:comp}
If a Lloyd-Topor program~$\Pi$ is $\Gamma$-tight, and an interpretation~$I$
satisfies both~$\Gamma$ and $\comp[\Pi]$, then $\gr_I(\Pi)$ is tight on $I^r$.
\end{lemma}

Finally, models of $\comp[\Pi]$ can be characterized in terms of satisfaction
and supportedness:

\begin{lemma}\label{lem:supported}
For any Lloyd-Topor program~$\Pi$,  an interpretation $I$ satisfies
$\comp[\Pi]$ iff~$I^r$ satisfies $\gr_I(\Pi)$ and is supported
by $\gr_I(\Pi)$.
\end{lemma}

Proofs of Lemmas \ref{lem:gen}--\ref{lem:supported} can be found in the
longer version of the paper, posted at
{\tt http://www.cs.utexas.edu/users/vl/papers/ltc-long.pdf}.

\section{Conclusion}

We proposed a new method for representing $\sm[F]$ in the language
of first-order logic.  It is more general than the approach of~\cite{fer09}.
Its relationship with the ideas of \cite{lee11b} requires further study.  This
method allows us, in particular, to prove the equivalence of some ASP
descriptions of dynamic domains to axiomatizations based on successor state
axioms.

The use of the stable model semantics of infinitary formulas \cite{tru12}
in the proof
of the main theorem illustrates the potential of that semantics as a tool for
the study of the operator~$\sm$.

\section*{Acknowledgements}

We are grateful to Joohyung Lee and to the anonymous referees for
useful comments.

\bibliographystyle{splncs}
\bibliography{bib}
\end{document}